\documentclass[aps]{revtex4}
\usepackage{graphicx}% Include figure files
\usepackage{amssymb}
\usepackage{amsfonts}

\newcommand{\mgcm}{mg~cm$^{-2}$}
\newcommand{\GT}{{\sc geant3}}
\newcommand{\GF}{{\sc geant4}}
\newcommand{\PEN}{{\sc penelope}}
\newcommand{\PENT}{{\sc penelope} (v.2003)}
\newcommand{\PENF}{{\sc penelope} (v.2005)}
\newcommand{\PENTR}{{\sc pen} 2003}
\newcommand{\PENFR}{{\sc pen} 2005}
\newcommand{\EG}{{{\sc egs}nrc}}
\newcommand{\MC}{{{\sc mcnpx}}}

\begin{document}

\title{Monte Carlo simulation of the electron transport through thin
slabs: \\A comparative study of \PEN, \GT, \GF, \EG \/ and \MC}

\author{M. Vilches$ ^{\,1}$, 
S. Garc\'{\i}a-Pareja$ ^{\,2}$, 
R. Guerrero$ ^{\,3}$, M. Anguiano$ ^{\,4}$ and
A.M. Lallena$ ^{\,4}$}

\affiliation{
$^{1}$Servicio de F\'{\i}sica y Protecci\'on Radiol\'ogica,
Hospital Regional Universitario ``Virgen de las Nieves'', Avda. de las
Fuerzas Armadas, 2, E-18014 Granada, Spain.\\
$^{2}$Servicio de Radiof\'{\i}sica Hospitalaria, Hospital
Regional Universitario ``Carlos Haya'', Avda. Carlos Haya, s/n,
E-29010 M\'alaga, Spain.\\
$^{3}$Servicio de Radiof\'{\i}sica, Hospital Universitario
``San Cecilio'', Avda. Dr. Ol\'oriz, 16, E-18012 Granada, Spain.\\
$^{4}$Departamento de F\'{\i}sica At\'omica, Molecular y
Nuclear, Universidad de Granada, E-18071 Granada, Spain.}

\begin{abstract}
The Monte Carlo simulation of the electron transport through thin
slabs is studied with five general purpose codes: \PEN, \GT, \GF, \EG
\/ and \MC. The different material foils analyzed in the old
experiments of Kulchitsky and Latyshev [Phys. Rev. 61 (1942) 254-266]
and Hanson {\it et al.} [Phys. Rev. 84 (1951) 634-637] are used to perform
the comparison between the Monte Carlo codes. Non-negligible
differences are observed in the angular distributions of the
transmitted electrons obtained with the some of the codes. 
The experimental data are reasonably well described by \EG,
\PENF \/ and \GF. A general good agreement is found for \EG \/ and
\GF \/ in all the cases analyzed.
\end{abstract}

%\pacs{25.30.Bf, 34.80.Bm}
\maketitle

\section{Introduction}

In Monte Carlo (MC) simulation of electron transport, the simulation
of all the interactions suffered by the electrons when they go through
a material, the so-called detailed simulation (DS), is, in general,
unpractical because of the long computing times required to reduce the
electron energies below the absorption threshold. Therefore, DS is
done in practice only if the energy of the electrons is low and/or the
targets are thin.

To solve the situation, ``condensed history'' schemes are employed in
much of the usual MC radiation transport codes of general purpose. In
this approach, a certain number of interactions are grouped and
described by means of a multiple scattering theory \cite{ber63}. In
condensed simulation (CS), electrons move in finite steps whose length
is calculated at the beginning of each step, using, for example, the
continuum slowing down approximation and a fixed average fractional
energy loss in the step.

An alternative approach is provided by the so-called mixed simulation
schemes, in which DS is used to simulate the ``hard'' interactions in
which the angular deflections and/or the energy losses are larger than
certain cut-off values, while the interactions which do not fulfill
these conditions, named ``soft'', are described within CS approach.

The statistical models used to determine the energy loss in CS provide
good enough results if the tracking steps are not too big. On the
contrary, the statistical treatment of the angular distribution linked
to the multiple scattering is a much more involved task. Since the
40's a considerable effort has been done to solve the problem and
various models (e.g. Goudsmit and Saunderson, Moli\`ere, Lewis) have
been developed \cite{sco63}.

Though multiple scattering theories can describe the angular
deflections produced by both elastic and inelastic collisions, it has
been a common practice to take care of the effects due to inelastic
collisions by correcting the distributions calculated considering only
elastic collisions \cite{ber88}. Only recently, Negreanu {\it et al.}
\cite{neg05} have treated both elastic and inelastic collisions on the
same footing, using accurate partial-wave differential cross sections.

The importance of the elastic scattering in the interaction of the
electrons with the materials and, in particular, in the dosimetry of
these particles, has been pointed out by different authors. Andreo
{\it et al.} \cite{and93} studied in detail the limitations of the
implementations of the Moli\`ere theory to be used in CS. Li and
Rogers \cite{li95} compared the {\sc egs}4 results with those obtained
by integrating analytically the Moli\`ere distribution, with those
quoted in ICRU Report 35 \cite{icru35} and with the experimental data
of Hanson {\it et al.} \cite{han51}. Urb\'an \cite{urb02} carried out a
comparison of the angular distributions obtained with \GF \/ with the
same data.

In the experimental side, the data available are very scarce. To the
best of our information, only Kulchitsky and Latyshev \cite{kul42} and
Hanson {\it et al.} \cite{han51} have performed measurements of the
multiple scattering of electrons in materials. Kulchitsky and Latyshev
\cite{kul42} studied the scattering of 2.25 MeV electrons by foils of
different materials from aluminum to lead. Their data were in good
agreement with the Goudsmit and Saunderson theory for the elements Al,
Fe, Cu, Mo, Ag and Sn, while for Ta, Au and Pb they found
non-negligible differences between theory and experiment. On the other
hand, Hanson {\it et al.} \cite{han51} measured the angular distributions
of 15.7 MeV incident electrons scattered by thin Be and Au foils. For
Au they found discrepancies with the predictions of the Goudsmit and
Saunderson theory and agreement with the calculations done according
to Moli\`ere theory. For Be the experimental data disagreed with the 
results obtained within the Moli\`ere approach.

In this work we analyze how different codes describe the scattering of
electrons by thin foils of different elements. Results for \GT, \GF,
\PEN, \EG \/ and \MC \/ are compared between them and with the few
experimental data available. In section 2 we give details concerning
the simulations we have performed and the MC codes which we have used
and which are relevant for the calculations performed. Also the way
how the data have been analyzed is discussed. In section 3 we describe
briefly the experiments of Kulchitsky and Latyshev \cite{kul42} and
Hanson {\it et al.} \cite{han51}. Section 4 is devoted to discuss the
results we have obtained: the calculations with different versions of 
some of the codes are discussed and the MC results are compared with the
experimental data of Hanson {\it et al.} \cite{han51} and Kulchitsky and
Latyshev \cite{kul42} and between themselves. In the last section we
draw our conclusions.

\section{Monte Carlo simulations}
\label{sec:MCS}

The simulations have been carried out using an elementary geometry
with a monoenergetic pencil beam impinging normally on a foil made of
a given material and with a given thickness. Both source and foil are
in vacuum. The incident beam defines the $z$ direction. The directions
of the (primary or secondary) electrons emerging from the foils define
the scattering angles which have been scored in a histogram for angles
between 0 and 45 degrees with 180 bins of 0.25 degrees. The values are
normalized to the solid angle unit given in stereoradian.

The statistical uncertainty in each bin is calculated as $\sigma =
\sqrt{\overline{Q}(1-\overline{Q})/N}$ with $\overline{Q}$ the value
per history in the bin and $N$ the number of histories
followed. Through the paper, these uncertainties are given as a number
between parentheses; e.g., 9.34(1) means 9.34$\pm$0.01.

\subsection{Monte Carlo codes}

As said before, we have performed calculations with various MC
codes. All these codes do DS for photons. Apart from the possible
differences in the photon cross sections, the main differences between
the codes are in the electron/positron transport. In what follows we
quote some details of the codes which are relevant for the MC
simulations done here.

\vspace{.2cm}
\centerline{\sl 1. \GT}
\vspace{.1cm}

\GT \/ \cite{geant3} is a system of detector description and
simulation tools designed for high-energy physics at CERN. It permits
the MC simulation of the transport of elementary particles and
different ions in elemental or compound materials for energies
ranging between 10 keV and 10 TeV.

The default model for the multiple scattering is based on the
Moli\`ere theory, but it is possible to select also the pure Gaussian
scattering according to the Rossi formulation \cite{ros41}. If the
number of Coulomb scatters is below 20, the Moli\`ere theory cannot be
applied and the simulation is performed within the plural scattering
regime in which the number of scatters is distributed according to a
Poisson distribution.

The tracking is controlled by means of a series of parameters. {\tt
DEEMAX} is the maximum fraction of kinetic energy which a particle can
lose in a step. This parameter as well as the multiple scattering
introduce upper limits to the step length. {\tt STEMAX} provides an
absolute upper limit to the step length, in cm, for each tracking
medium. {\tt STMIN} imposes a lower limit to this step length, also in
cm, which permits to accelerate the stopping of those particles with
very small energies. {\tt EPSIL} determines the boundary crossing
precision in cm.  In \GT \/ one can use an automatic mode for the
calculation of these parameter. This is done by selecting {\tt
AUTO}=1. Also, it is possible to fix energy cuts for the different
particles. In the calculations we have done, these cuts have been
fixed to 10 keV for both photons and electrons.

In this work we have used the version 3.2114 of \GT. The simulations
have been done using the automatic mode. To check the feasibility of
the corresponding results, we have performed simulations for the Au
foil of 18.66 \mgcm \/ with eight different sets of the parameters
{\tt STEMAX}, {\tt DEEMAX} and {\tt STMIN}, varying from 0.001 to 0.1,
the first two, and from 0.0001 to 0.01, the last one. Despite the fact
that this is the most exigent case, because correspond to the smaller
number of interactions, the results obtained are compatible (within
the statistical uncertainty) with those provide by the {\tt AUTO=1}
option.

\vspace{.2cm}
\centerline{\sl 2. \GF}
\vspace{.1cm}

\GF \/ code \cite{ago03} is an object-oriented C++ toolkit which
permits the MC simulation of the radiation transport in material media
for a great variety of particles, materials and energies. For the
electromagnetic interactions of photons and electrons, \GF \/ permits
to use three different physics models: {\it Standard}, {\it
Low-energy} and {\it Penelope}. In this work the package {\it
Low-energy} has been used.

The multiple scattering approach used in \GF \/ has been developed by
Urb\'an \cite{urb02} based on the Lewis theory \cite{sco63}. The
energy loss is calculated from the actual path length which is
computed in every step after performing a path length correction. In
addition, a lateral displacement and a scattering angle are sampled
from given distributions.

In \GF, particles are produced if their energies are above given
thresholds which are specified in terms of distances, for each volume
in the geometry, and internally converted to energy. For the {\it
Low-energy model}, the minimum value for these thresholds is 250
eV. In our calculations we have fixed 10 keV for electrons and 1 keV 
for photons.

Particles are followed until their kinetic energy is zero, but
tracking cuts can be fixed. Other parameters (some of them not
described in the \GF \/ manuals \cite{G4phys,G4deve}) provide
additional control of electron step. By means of the {\it
G4UserLimits} class it is possible to fix a maximum step size and a
maximum track length, as well as the tracking cuts above
mentioned. The variable {\tt dRoverRange} determines the maximum
fraction of the stopping range that can be travelled by an electron in
a step. The parameter {\tt finalRange} fix the minimum step size below
which the electron is absorbed locally. Finally, the {\tt fr} variable
permits to control the step size when the electron is transported away
from a boundary into a new volume. Defaults are {\tt dRoverRange}=0.2,
{\tt finalRange}=0.2~mm, {\tt fr}=0.02 and no maximum step size. These
are the values we have used in our simulations.

We have performed test simulations using {\tt dRoverRange}=0.05, {\tt
finalRange}=1~nm, {\tt fr}=0.02 and no maximum step size. No
significant differences have been found with respect to the
simulations corresponding to the default values.

We have used the versions 8.0 (patch01) of \GF \/ and {\sc
g4emlow3.0} of the {\it Low-energy} package. Some additional
simulations with the versions 4.1 and 6.0 have also been performed for
comparison.

\vspace{.2cm}
\centerline{\sl 2. \PEN}
\vspace{.1cm}

\PEN \/ \cite{sal03} is a general purpose MC code which performs
simulations of coupled electron-photon transport. It can be applied
for energies ranging from a few hundred eV up to 1~GeV and for
arbitrary materials. Besides, \PEN \/ permits a good description of
the particle transport at the interfaces and presents a more accurate
description of the electron transport at low energies in comparison to
other general purpose MC codes. Details about the physical processes
considered can be found in \cite{sal03}.

In \PEN \/ electrons and positrons are simulated by means of a mixed
scheme where, as said above, collisions are classified as ``hard'' and
``soft''. The electron tracking is controlled by means of four
parameters. $C_1$ and $C_2$ refer to elastic collisions. $C_1$ gives
the average angular deflection due to a elastic hard collision and to
the soft collisions previous to it. $C_2$ represents the maximum value
permitted for the average fractional energy loss in a step. On the
other hand, $W_{\rm cc}$ and $W_{\rm cr}$ are energy cutoffs to
distinguish hard and soft events. Thus, the inelastic electron
collisions with energy loss $W<W_{\rm cc}$ and the emission of
bremsstrahlung photons with energy $W<W_{\rm cr}$ are considered in
the simulation as soft interactions. The maximum step size can be
controlled using the parameter $s_{\rm max}$.

We have used two version of this code: 2003 and 2005. In what refers
to the multiple scattering the main difference between both versions
is that elastic electron collisions are simulated by means of a model
based on the Wentzel angular distribution \cite{sal03} in the version
2003, while in the 2005 these collisions are simulated by using
relativistic (Dirac) partial-wave differential cross sections
generated by using the computer code {\sc elsepa} \cite{sal05}.

Some of the simulations performed with \PEN \/ have been done with a 
set of parameters which have been used in different simulations in 
which thin slabs are present (see e.g. \cite{sem01}). We label these
simulations as ``safe'' and the values of the parameters used are: 
$W_{\rm cc}=5$~keV, $W_{\rm cr}=1$~keV, $C_1=C_2=0.05$. Photons were 
simulated down to 10~keV. Electrons and positrons were absorbed when 
they slow down to kinetic energies of 100~keV. The mixed scheme of 
the simulation in \PEN \/ permits to perform fully ``detailed'' 
simulations. We have done calculations also in this approach by 
selecting $W_{\rm cc}=0$, $W_{\rm cr}=-1000$~keV, $C_1=C_2=0$. 
In this case the absorption energies were fixed to 100 eV for all 
particles. In all the simulations done with \PEN \/ $s_{\rm max}$ 
was taken to be 1/20 of the width of the foil, as it is recommended 
in \cite{sal03}.

\vspace{.2cm}
\centerline{\sl 2. \EG}
\vspace{.1cm}

\EG\/ \cite{kaw03} is a general purpose package designed for the
Monte Carlo simulation of the coupled transport of electrons and
photons in arbitrary geometries. Particle energies
above a few keV up to several hundreds of GeV can be considered.

Multiple scattering of charged particles is described by means of an
approach developed by Kawrakow and Bielajew \cite{kaw98a} in which
most of the shortcomings of the Moli\`ere multiple scattering theory
were fixed. Within this approach, track steps are simulated using a
single scattering model for short steps and a multiple scattering
model for large steps. In addition one can select between Rutherford
scattering or scattering including both relativistic and spin
effects. The electron transport algorithm is due to Kawrakow and
Bielajew \cite{kaw98b} and it is usually known as {\sc presta-ii}.

In \EG\/ the simulation is controlled by the following parameters:
{\tt SMAXIR}, which defines upper limits on the step size in each
region in the geometry; {\tt ESTEPR}, which fixes the maximum
fractional energy loss per electron step in each region; {\tt ESTEPE},
which is a global energy loss constraint, and {\tt XIMAX}, which gives
the maximum first Goudsmit and Saunderson moment per step. In our
simulations, we have used {\sc presta-ii} (which is the default
transport algorithm) and, as indicated in the manual \cite{kaw98a},
the default values {\tt SMAXIR}=10$^{10}$~cm, {\tt ESTEPR}=1, {\tt
ESTEPE}=0.25 and {\tt XIMAX}=0.5 must not be changed. The version
V4-r2-2-3 has been used for calculations. Photons and electrons were
followed down to 10 keV.

\vspace{.2cm}
\centerline{\sl 2. \MC}
\vspace{.1cm}

\MC\/ \cite{mcnpx-manual} is a general purpose MC code which permits
the description of the transport of different particles in arbitrary
materials. Photons, electrons and neutrons, as well as other 29
particles between leptons, baryons, mesons and even light ions can be
considered. The upper energy limits for electrons and photons are 1
and 100 GeV, respectively. A lower limit of 1 keV is fixed for these
particles.

The angular deflections in the multiple scattering of electrons are
calculated according to the Goudsmit and Saunderson theory. When
electron energies are below 0.256 MeV, the corresponding cross
sections are obtained from numerical tabulations developed by Riley
{\it et al.} \cite{ril75}. For higher energy electrons, the cross sections
are approximated as a combination of the Mott and Rutherford cross
sections, including a correction factor which takes care of the
screening.

The particle tracking is governed by {\tt EMCPF}, which is the upper 
energy limit for detailed photon physics treatment, {\tt EMAX}, which 
fixes the upper limit for electron energy, and the low energy cutoffs.
We have used the default value (100 MeV) for {\tt EMCPF} and 
{\tt EMAX=20} MeV, while the low energy cutoffs were fixed to 1 keV.
In our simulations we have used the version 2.4.0.

In the calculations done with \MC, the angular distributions obtained
by using the binning indicated in Sect. \ref{sec:MCS} showed a
stepwise shape in which various neighbour bins had the same value.
This is due to the fact that the Goudsmit and Saunderson angular
distribution is sampled by using the inverse transform method applied
to a histogram of 34 bins with widths equal or larger than 1
degree. In the cases analyzed here, where few interactions occur, this
lack of precision in the sampling produces wrong results. To solve
this problem, the bins in the simulations performed with \MC\/ have
been enlarged to 1 degree.

\section{Description of the experiments}

In the two experiments mentioned above, the angular distribution of 
electrons scattered by foils of different materials and widths were 
measured. 

Kulchitsky and Latyshev \cite{kul42} studied the scattering of 2.25
MeV electrons by foils of Al, Fe, Cu, Mo, Ag, Sn, Ta, Au and Pb. The
thicknesses (see first row in Table \ref{tab:EXPER-K}) were chosen in
order to maintain the half-width of the distribution $\sim$10 degrees
in all cases, thus ensuring the validity of the multiple scattering
theory. Measurements for angles between 0 and 35-40 degrees were
done. Electrons with energies in the range 2250$\pm$23 keV formed
their beam.

\begin{table}[b]
\begin{center}
\begin{tabular}{ccccccccc}
\hline\hline
& Al & Fe & Cu & Mo & Ag & Ta & Au & Pb \\ \cline{2-9}
thickness [\mgcm] & 
 26.6 & 15.4 & 17.15 & 12.4 & 11.55 & 8.9 & 8.9 & 7.9 \\ \hline\hline
$\theta^{\rm Kul}_{1/e}$ [deg] & 
9.50 & 9.60 & 10.40 & 10.25 & 10.20 & 9.85 & 9.20 & 9.70 \\
$\theta^{\rm exp}_{1/e}$ [deg] & 
9.49 & 9.33 & 10.61 & 9.97 & 10.21 & 9.91 & 10.07 & 9.36 \\
\hline\hline
\end{tabular}
\end{center}
\vspace*{-.2cm}
\caption{\small Comparison of the values of the parameter $\theta^{\rm
exp}_{1/e}$ obtained after reanalyzing the experimental data of
Kulchitsky and Latyshev \cite{kul42} with those quoted by these
authors. The thicknesses of the different foils are also given.
\label{tab:EXPER-K}}
\end{table}

The data obtained were compared to the Goudsmit and Saunderson
multiple scattering theory predictions. Specifically, the comparison
was done for the angle $\theta_{1/e}$ at which the angular
distribution reduces by a factor $1/e$. Kulchitsky and Latyshev
estimated the uncertainty in $\theta_{1/e}$ as 3-4\% at most, but they
provided only relative values of the distributions measured, what
impedes absolute comparisons with our calculations.

Hanson {\it et al.} \cite{han51} measured the angular distributions of
electrons scattered by thin Be (257 and 495 \mgcm) and Au (18.66 and
37.28 \mgcm) foils for 15.7 MeV incident electrons. In the case of Be
foils, the measurements were done up to 6 degrees. For Au foils the
angular range was extended up to 30 degrees. The beams included
electrons with energies within 6 percent of the maximum energy.

Hanson {\it et al.} compared their experimental data to the
predictions of the Goudsmit and Saunderson and the Moli\`ere theories
for the angle $\theta_{1/e}$. They do not quote the uncertainties of
their measurements but provide absolute values of the distributions,
what allows a more complete comparison.

The results obtained with the different MC codes have been compared to
the experimental data by means of the quantities $\theta_{1/e}$ and,
when possible, $F(0)$. This last corresponds to the maximum of the
angular distribution, at $\theta=0$. To do that we have fitted a Gaussian
function
\begin{equation}
F(\theta) \, = \, \displaystyle F(0) \, \exp{
\left[ \displaystyle -\frac{\theta^2}{(\theta_{1/e})^2} \right] }
\end{equation}
to the simulated angular distributions, in the angular range between 0
and the first angle larger than $\theta_{1/e}$. The Levenberg-Marquadt
method \cite{recipes} has been used.

\begin{table}
\begin{center}
\begin{tabular}{ccccc}
\hline\hline
& $\theta^{\rm Han}_{1/e}$ [deg] & $\theta^{\rm exp}_{1/e}$ [deg] & 
$F^{\rm Han}(0)$ [sr$^{-1}$] & $F^{\rm exp}(0)$ [sr$^{-1}$] \\ \hline
  Be (257 \mgcm) & 3.06 &      & 104.06 &  \\
  Be (495 \mgcm) & 4.25 &      &  50.23 &  \\
Au (18.66 \mgcm) & 2.58 & 2.50 & 144.77 & 143.29  \\
Au (37.28 \mgcm) & 3.76 & 3.71 &  65.66 &  64.68  \\
\hline\hline
\end{tabular}
\end{center}
\vspace*{-.2cm}
\caption{\small Comparison of the values of the parameters
$\theta^{\rm exp}_{1/e}$ and $F^{\rm exp}(0)$ obtained after
reanalyzing the experimental data of Hanson {\it et al.} \cite{han51} with
those quoted by these authors. We have not calculated the values for
the Be foils because the angular distributions are not available in
Ref. \cite{han51}.
\label{tab:EXPER-H}}
\end{table}

The experimental data of Kulchitsky and Latyshev \cite{kul42} and
Hanson {\it et al.} \cite{han51} have been reanalyzed following the same
criteria used for the simulated distributions. In this way we have
recalculated the ``experimental'' $\theta_{1/e}$ and $F(0)$. We call
the values obtained in this way $\theta^{\rm exp}_{1/e}$ and $F^{\rm
exp}(0)$, respectively. We give them in Tables \ref{tab:EXPER-K} and
\ref{tab:EXPER-H} where they are compared to the values quoted in
Refs. \cite{kul42}, $\theta^{\rm Kul}_{1/e}$, and \cite{han51},
$\theta^{\rm Han}_{1/e}$ and $F^{\rm Han}(0)$. The values $F^{\rm
exp}(0)$ have not been calculated for the experiment of Kulchitsky and
Latyshev because the absolute distributions are not available in
\cite{kul42}. On the other hand, the values $\theta^{\rm exp}_{1/e}$
and $F^{\rm exp}(0)$ have not been calculated for the Be foils in the
experiment of Hanson {\it et al.} because they do not provide the
corresponding angular distributions.

In the case of the experiment of Kulchitsky and Latyshev (see Table
\ref{tab:EXPER-K}), the maximum difference between our results and
those quoted in Ref. \cite{kul42} is for Au foil (8.6\%). In the case
of the Au foils in the experiment of Hanson {\it et al.}, the maximum
difference is smaller than 3.5\% (see Table \ref{tab:EXPER-H}).

To finish we mention here the fact that there is a mistake concerning
the Sn data in the experiment of Kulchitsky and Latyshev. The
thickness quoted in Table 1 of Ref. \cite{kul42} is incompatible with
the value $\theta_{1/e}=$10.9 degrees they obtained. All the
simulations we have performed disagree by more than 20\% with this
value. This disagreement can be seen also in Fig. 5 of
Ref. \cite{neg05} where the data were compared to calculations using
the Lewis multiple scattering theory. We have not considered this
foil in our discussion.

\section{Results}

\subsection{Comparison of different code versions}

In the case of \GF \/ and \PEN, different versions have been considered
in the calculations. Here we compare the results obtained between them.

\begin{figure}[htb]
\begin{center}
\parbox[c]{15cm}
{\includegraphics[scale=0.6]{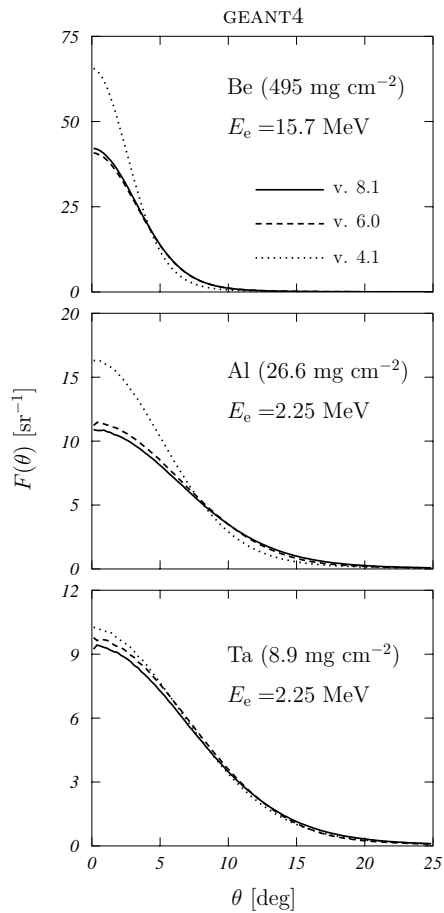}}
\end{center}
\vspace*{-.5cm}
\caption{\small Angular distributions obtained with the versions 8.0
(solid curves), 6.0 (dashed curves) and 4.1 (dotted curves) of the \GF
\/ code. We show the results for 15.7 MeV electrons impinging in a Be
foil with 495 \mgcm \/ (upper panel) and for 2.25 MeV electrons
incident on an Al (medium panel) and Ta (lower panel) foils with 26.6
and with 8.9 \mgcm, respectively. Statistical uncertainties
(1$\sigma$) are smaller than 2.5\% and have not been plotted.
\label{fig:comp-G4}
}
\end{figure}

\begin{table}
\begin{center}
\begin{tabular}{ccccc}
\hline\hline
 & & {4.1} & {6.0} & {8.0} \\
\hline\hline
Be           & $F(0)$ [sr$^{-1}$]   & 66.13(8) & 40.63(4) & 41.84(5) \\
(495 \mgcm)  & $\theta_{1/e}$ [deg] & 3.762(4) & 4.778(5) & 4.735(5)  \\ \hline
Al           & $F(0)$ [sr$^{-1}$]   & 16.23(2) & 11.44(1) & 10.83(1) \\
(26.6 \mgcm) & $\theta_{1/e}$ [deg] & 7.473(9) &  9.17(1) &  9.33(1) \\ \hline
Ta           & $F(0)$ [sr$^{-1}$]   & 10.13(1) &  9.72(1) &  9.34(1)  \\
(8.9 \mgcm)  & $\theta_{1/e}$ [deg] &  9.51(1) & 10.03(1) & 10.04(1)\\ 
\hline\hline
\end{tabular}
\end{center}
\vspace*{-.2cm}
\caption{\small Values of the parameters $F(0)$ and $\theta_{1/e}$
obtained for the calculations shown in Fig. \ref{fig:comp-G4} and
performed with the three versions of \GF \/ considered in this work.
The uncertainties are given at 1$\sigma$ level.
\label{tab:comp-G4}}
\end{table}

In Fig. \ref{fig:comp-G4} we show the angular distributions obtained
with the versions 4.1, 6.0 and 8.0 of \GF \/ for three different
cases: 15.7 MeV electrons impinging on a Be foil with 257 \mgcm (upper
panel) and 2.25 MeV electrons incident on Al (medium panel) and Ta
(lower panel) foils with 26.6 and 8.9 \mgcm, respectively. As we can
see, the calculations performed with the versions 6.0 and 8.0 provide
rather similar results, while those found for the 4.1 version disagree
with them. The differences are larger the smaller the atomic number of
the foil material is. In order to quantify these differences we show,
in Table \ref{tab:comp-G4} the values of the parameters $F(0)$ and
$\theta_{1/e}$ corresponding to these angular distributions. For Be
and Al, $\theta_{1/e}$ is 25\% larger for the version 8.0 than for the
4.1 one, while for Ta is less than 6\%. On the other hand, for Be and
Al, the $F(0)$ value obtained with the version 4.1 is more than 50\%
larger than for the version 8.0, while is only 8\% larger for Ta. From
these results it is evident that the various changes in the details
concerning the multiple scattering theory carried out from version 4.1
till version 8.0 (see \cite{urb02}) are not at all negligible.

\begin{figure}[htb]
\begin{center}
\parbox[c]{15cm}
{\includegraphics[scale=0.6]{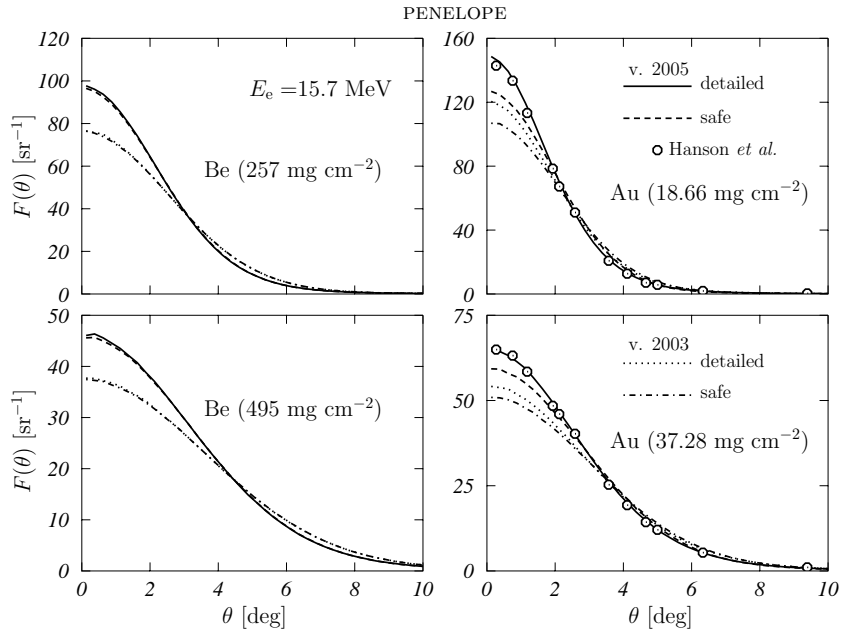}}
\end{center}
\vspace*{-.5cm}
\caption{\small Angular distributions obtained with \PEN \/ for the
four foils of the experiment of Hanson {\it et al.} \cite{han51}. Detailed
simulations with the versions 2003 (dotted curves) and 2005 (solid
curves) and simulations performed with the ``safe'' parameter
indicated in the text also with versions 2003 (dashed-dotted curves)
and 2005 (dashed curves) are shown. For the two Au foils, the open
circles represent the experimental data. Statistical uncertainties
(1$\sigma$) are smaller than 2\% and have not been plotted.
\label{fig:comp-PE}
}
\end{figure}

\begin{table}
\begin{center}
\begin{tabular}{cccccc}
\hline\hline
& & \multicolumn{2}{c}{2003} & \multicolumn{2}{c}{2005} \\
\cline{3-4} \cline{5-6}
& & ``safe'' & detailed & ``safe'' & detailed \\\hline\hline
           Be & $F(0)$ [sr$^{-1}$]   & 75.9(1)  & 76.5(1)  & 95.8(2)  & 97.0(2)  \\
  (257 \mgcm) & $\theta_{1/e}$ [deg] & 3.640(6) & 3.620(6) & 3.173(5) & 3.141(5) \\ \hline
           Be & $F(0)$ [sr$^{-1}$]   & 37.51(8) & 37.73(8) & 45.7(1)  & 46.1(1)  \\
   (495 \mgcm)& $\theta_{1/e}$ [deg] & 5.18(1)  & 5.15(1)  & 4.595(9) & 4.563(9) \\ \hline
           Au & $F(0)$ [sr$^{-1}$]   & 107.8(2) & 119.3(3) & 126.6(3) & 146.9(3) \\
(18.66 \mgcm) & $\theta_{1/e}$ [deg] & 3.033(5) & 2.796(5) & 2.791(5) & 2.480(4) \\ \hline
           Au & $F(0)$ [sr$^{-1}$]   & 51.0(1)  & 53.9(1)  & 59.4(1)  & 64.4(1)  \\
(37.28 \mgcm) & $\theta_{1/e}$ [deg] & 4.376(8) & 4.181(8) & 4.032(8) & 3.763(7) \\ 
\hline\hline
\end{tabular}
\end{center}
\vspace*{-.2cm}
\caption{\small Values of the parameters $F(0)$ and $\theta_{1/e}$
obtained for the calculations shown in Fig. \ref{fig:comp-PE} and
performed with the two versions of \PEN \/ considered in this work.
The uncertainties are given at 1$\sigma$ level.
\label{tab:comp-PE}
}
\end{table}

For \PEN, the versions 2003 and 2005 have been considered and for each
one, both ``detailed'' and ``safe'' simulations have been
performed. In Fig. \ref{fig:comp-PE} we show the corresponding angular
distributions. Full and dotted lines represent the results of detailed
calculations with versions 2005 and 2003 respectively. Dashed and
dashed-dotted curves correspond to the ``safe'' simulations.  The
values of the characteristic parameters are given in Table
\ref{tab:comp-PE}. Two findings deserve a comment. First, the results
obtained with the ``safe'' simulations are very similar to those
obtained with the ``detailed'' simulation for the two Be foils (left
panels). In these cases the differences are 2\% at most. However, for
the Au slabs, the differences between the results obtained in both
simulations are larger. For $F(0)$, these differences are above 20\%
for the version 2003 and around 15\% for the 2005. The differences
between the values obtained for $\theta_{1/e}$ within ``safe'' and
detailed simulations are around 5\% for the version 2003 and slightly
larger for the version 2005. The discrepancies observed between
``safe'' and ``detailed'' simulations are linked to the number of
interactions suffered by the electrons in the foils. In case of Be,
electrons suffer $\sim 10$ hard interactions in average; for Au foils,
the average number of hard interactions is smaller than 1. As
indicated in \cite{sal03}, this makes ``safe'' simulations to be into
agreement with the ``detailed'' ones for Be foils and to show the
differences discussed in the case of the Au foils. By fixing
adequately the tracking parameters these differences can be strongly
reduced. For example, in the case of the thin Au foil, a reduction of
the parameter $C_1$ to 0.01 makes the difference at $\theta=0$ degrees
to diminish from 14.7 to 6.7\%.

The second point to be noted concerns the differences observed between
the two versions of \PEN. By comparing the results corresponding to
the detailed simulations, differences between 19 and 27\% (for $F(0)$)
and 11 and 15\% (for $\theta_{1/e}$) are found, the larger differences
occurring for the thin Be slab and the smaller for the thick Au foil. 

In addition, the substitution of the modified Wentzel angular
distribution (used in the version 2003) by the relativistic
partial-wave differential cross sections (which incorporates the
version 2005) permits, in the case of the detailed simulations, a very
good description of the experimental data for the Au foils (which are
plotted in Fig. \ref{fig:comp-PE} with open circles).

\begin{figure}[hb]
\begin{center}
\parbox[c]{14cm}
{\includegraphics[scale=0.5]{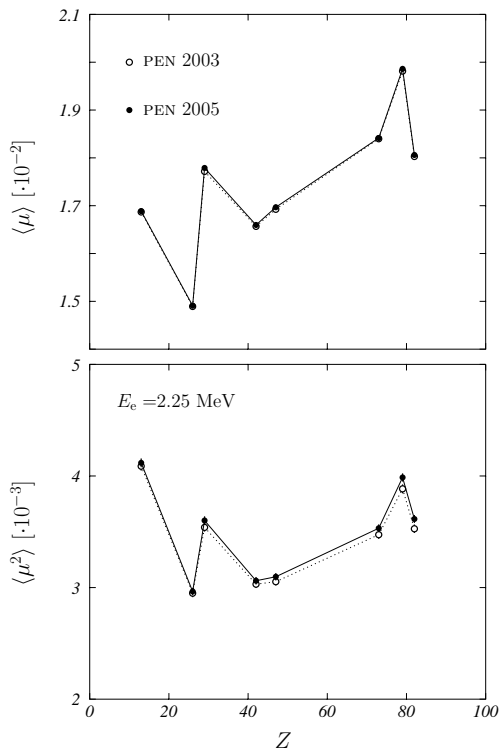}}
\end{center}
\vspace*{-.5cm}
\caption{\small Values of $\langle \mu \rangle$ (upper panel) and
$\langle \mu^2 \rangle$ (lower panel) as a function of the atomic
number $Z$ of the foil materials in the experiment of Kulchitsky and
Latyshev \cite{kul42}. Detailed simulations with the versions 2003
(open circles) and 2005 (black points) of \PEN \/ are
shown. Statistical uncertainties (1$\sigma$) are smaller than the
symbols used to plot the results.
\label{fig:comp-THEMEAN-PE}
}
\end{figure}

In order to go deeper into the differences observed between the two
versions of \PEN, we have calculated the mean values $\langle \mu
\rangle$ and $\langle \mu^2 \rangle$ of the quantity $\mu=(1-\cos
\theta)/2$ which is commonly used to measure polar angular deflections
produced by single scattering events, instead of the scattering angle
$\theta$. The results of the calculations for the material foils
considered in the experiment of Kulchitsky and Latyshev \cite{kul42}
are shown in Fig. \ref{fig:comp-THEMEAN-PE}. In these calculations
both the forward ($0 \leq \theta \leq 90$~deg) and the backward
($90~{\rm deg} \leq \theta \leq 180$~deg) distributions have been
considered to perform the corresponding integrals. It should be
pointed out the almost perfect agreement between the results obtained
with the two versions of the code. This was expected according to the
fact that, as indicated in the \PENT \/ manual \cite{sal03}, the
multiple scattering theory used in this version (the modified Wentzel
model) was fixed to reproduce the values of $\langle \mu \rangle$ and
$\langle \mu^2 \rangle$ obtained for the actual partial-wave
differential cross section which are used, instead, in \PENF. In
principle, this equality ensures a reasonable description of the
multiple scattering in ``normal'' simulation conditions with enough
interactions. It is clear that in the cases analyzed here (where
electrons go through rather thin foils) the differences between the
models used show up.

In what follows we have considered only the version 8.0 of \GF \/ and
the detailed simulations of the versions 2003 and 2005 of \PEN.

\subsection{Comparison with experiment}

\begin{figure}[htb]
\begin{center}
\parbox[c]{15cm}
{\includegraphics[scale=0.6]{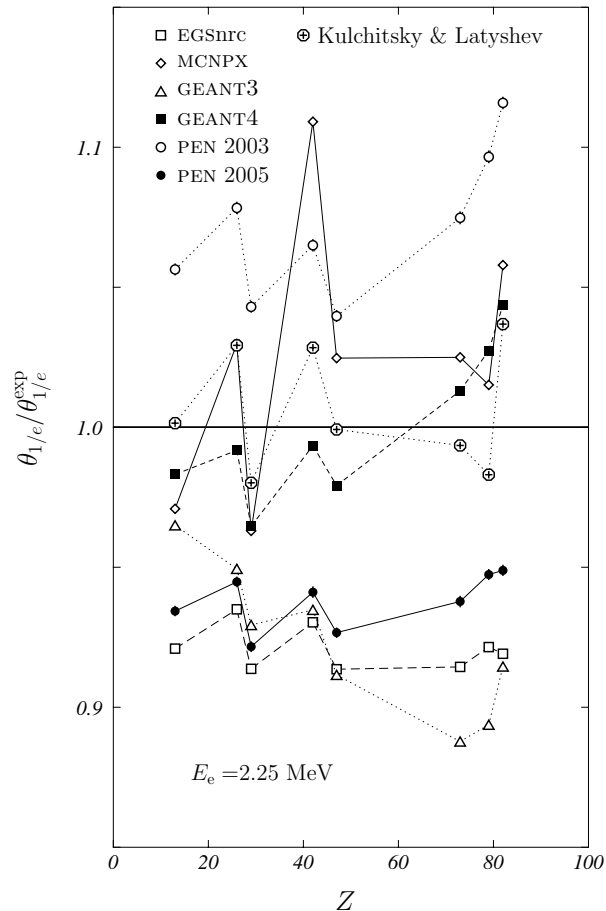}}
\end{center}
\vspace*{-.5cm}
\caption{\small Ratio of the values $\theta_{1/e}$ obtained with the
various MC codes to the values $\theta^{\rm exp}_{1/e}$ obtained in
our reanalysis of the corresponding experimental data as a function of
the atomic number $Z$ of the foil materials in the experiment of
Kulchitsky and Latyshev \cite{kul42}. Circled plus symbols correspond
to the ratio between the values $\theta^{\rm Kul}_{1/e}$ quoted by
Kulchitsky and Latyshev (see Table \ref{tab:EXPER-K}) and $\theta^{\rm
exp}_{1/e}$. The lines between symbols are only to guide
eyes. Uncertainties are considered at the 1$\sigma$ level and in most
cases are smaller than the symbols used to represent the data.
\label{fig:comp-KUL}
}
\end{figure}

First we have compared the results of the simulations to the
experimental data of Kulchitsky and Latyshev \cite{kul42}. In
Fig. \ref{fig:comp-KUL} we show the ratio of the values $\theta_{1/e}$,
obtained with the various MC codes, to the values $\theta^{\rm
exp}_{1/e}$, obtained in our reanalysis of the corresponding
experimental data, as a function of the atomic number $Z$ of the foil
materials. In addition, the ratio $\theta^{\rm Kul}_{1/e}/\theta^{\rm
exp}_{1/e}$ is plotted as reference (see circled plus symbols.)

\begin{table}
\begin{center}
\begin{tabular}{ccccccc}
\hline\hline
& \EG & \MC & \GT & \GF & \PENTR & \PENFR \\ 
\hline\hline
 & -8.7\% (Ag) & 10.9\% (Mo) & -11.2\% (Ta) & 4.4\% (Pb)  & 11.6\% (Pb)& -7.8\% (Cu) \\
 & -6.5\% (Fe) & -3.7\% (Cu) & -3.5\% (Al)  & -3.5\% (Cu) & 4\% (Ag)   & -5.1\% (Pb) \\
\hline
$S$ [deg$^2$] & 4.9 & 1.9 & 5.2 & 0.5 & 4.3 & 3.1 \\
\hline\hline
\end{tabular}
\end{center}
\vspace*{-.2cm}
\caption{\small Relative differences 
$(\theta_{1/e}-\theta^{\rm exp}_{1/e})/\theta^{\rm exp}_{1/e}$ 
with maximum (first row) and minimum (second row) absolute values, 
obtained for each MC code for the foils in 
the experiment of Kulchitsky and Latyshev \cite{kul42}. The corresponding foil 
materials are given. Last row shows the values of $S$ as defined by Eq. (\ref{eq:sum2}).
\label{tab:KU-results}
}
\end{table}

While \EG, \GT \/ and \PENF \/ underestimate in all cases the values
of $\theta^{\rm exp}_{1/e}$, \PENT \/ overestimates it for all foils
and \MC \/ and \GF \/ show, in this respect, a $Z$ dependent behaviour. 
The differences between the values $\theta_{1/e}$ obtained for the 
different codes and those we have found in our reanalysis of the 
experimental data, which show the maximum and minimum absolute values,
are shown in Table \ref{tab:KU-results}. As we can see, \EG \/ and 
\PENF \/ maintain these differences within a range of 3\%, \GT, \GF \/ 
and \PENT, within 8\%, while \MC \/ shows a range bigger than 14\%.
 
In order to quantify globally these differences we have calculated, 
for each MC code, the sum
\begin{equation}
S \, = \, \sum_i \, 
\left( \theta^i_{1/e} - \theta^{\rm exp}_{1/e} \right)^2 \, ,
\label{eq:sum2}
\end{equation}
where $i$ runs over the eight different foils considered. The better
codes correspond to those with the smaller values of $S$. The results
obtained are shown in the last row of Table \ref{tab:KU-results}. On
the base of this index, the ``best'' results are those obtained with
\GF \/ while the large discrepancies are obtained with \GT, \EG \/ and
\PENT.

\begin{figure}[htb]
\begin{center}
\parbox[c]{15cm}
{\includegraphics[scale=0.6]{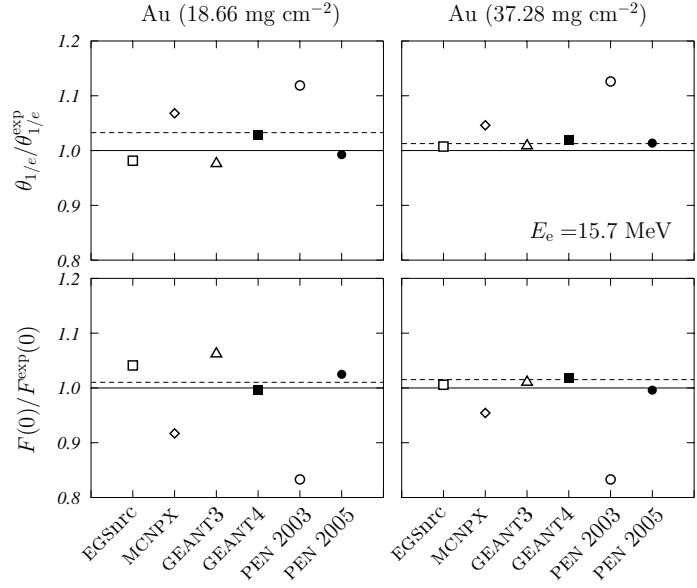}}
\end{center}
\vspace*{-.5cm}
\caption{\small Ratio of the values $\theta_{1/e}$ (upper panels) and
$F(0)$ (lower panels) obtained with the various MC codes to the values
$\theta^{\rm exp}_{1/e}$ and $F^{\rm exp}(0)$, respectively, obtained
in our reanalysis of the data for the two Au foils in the experiment of
Hanson {\it et al.} \cite{han51}. The dashed lines represent the ratios
between the values $\theta^{\rm Han}_{1/e}$ and $F^{\rm Han}(0)$
quoted by Hanson {\it et al.} (see Table \ref{tab:EXPER-H}) and $\theta^{\rm
exp}_{1/e}$ or $F^{\rm exp}$. Uncertainties are considered at the
1$\sigma$ level and are smaller than the symbols used to represent the
data.
\label{fig:comp-AU}
}
\end{figure}

\begin{figure}[htb]
\begin{center}
\parbox[c]{15cm}
{\includegraphics[scale=0.6]{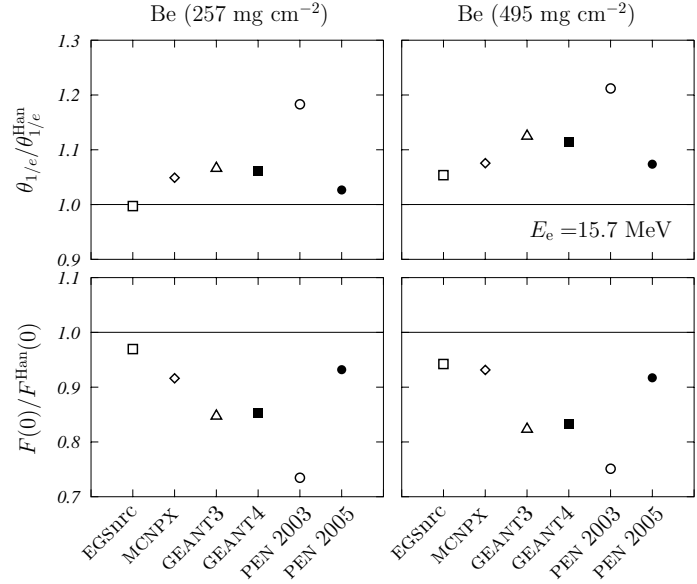}}
\end{center}
\vspace*{-.5cm}
\caption{\small Ratio of the values $\theta_{1/e}$ (upper panels) and
$F(0)$ (lower panels) obtained with the various MC codes to the values
$\theta^{\rm Han}_{1/e}$ and $F^{\rm Han}(0)$, respectively, quoted
for the Be foils in the experiment of Hanson {\it et al.} \cite{han51}
(see Table \ref{tab:EXPER-H}). Uncertainties are considered at the
1$\sigma$ level and are smaller than the symbols used to represent the
data.
\label{fig:comp-BE}
}
\end{figure}

A slightly different situation is found when we compare the results of
our simulations to the experimental data of Hanson {\it et al.}
\cite{han51}. In Fig. \ref{fig:comp-AU} we show the ratio of the
values $\theta_{1/e}$ (upper panels) and $F(0)$ (lower panels)
obtained with the various MC codes to the values $\theta^{\rm
exp}_{1/e}$ and $F^{\rm exp}(0)$, respectively, for the two Au foils
in the experiment of Hanson {\it et al.} \cite{han51}. The dashed lines
represent the ratios between the values $\theta^{\rm Han}_{1/e}$ and
$F^{\rm Han}(0)$ quoted by Hanson {\it et al.} (see Table
\ref{tab:EXPER-H}) and $\theta^{\rm exp}_{1/e}$ or $F^{\rm exp}$. As
we can see, \EG, \GT, \GF \/ and \PENF \/ provide a very good
description of the experiment, \MC \/ slightly overestimates
(underestimates) $\theta^{\rm exp}_{1/e}$ ($F^{\rm exp}(0)$) and \PENT
\/ gives the ``worst'' results.  The excellent agreement found for \GF
\/ is not surprising because the multiple scattering theory developed
by Urb\'an and which is used in this code has been tuned to reproduce
these Au experimental data \cite{urb02}.

Fig. \ref{fig:comp-BE} shows the comparison with the experimental
results of Hanson {\it et al.} \cite{han51} for the two Be foils. As said
before, data are not available for these foils and we could not
perform the corresponding reanalysis. In this case we have used
directly the values $\theta^{\rm Han}_{1/e}$ and $F^{\rm Han}(0)$
quoted by Hanson {\it et al.} (see Table \ref{tab:EXPER-H}) as reference
values. Here the ``best'' description of the experiment is provided by
\EG, \MC \/ and \PENF. \GT \/ and \GF \/ show very similar results and
the major differences occur for \PENT, reaching values around 20\% or
even larger, both for $\theta_{1/e}$ and $F(0)$. On the other hand,
and as in the case of the Au slabs, relative results are rather
similar for both foils.

\begin{figure}[htb]
\begin{center}
\parbox[c]{15cm}
{\includegraphics[scale=0.6]{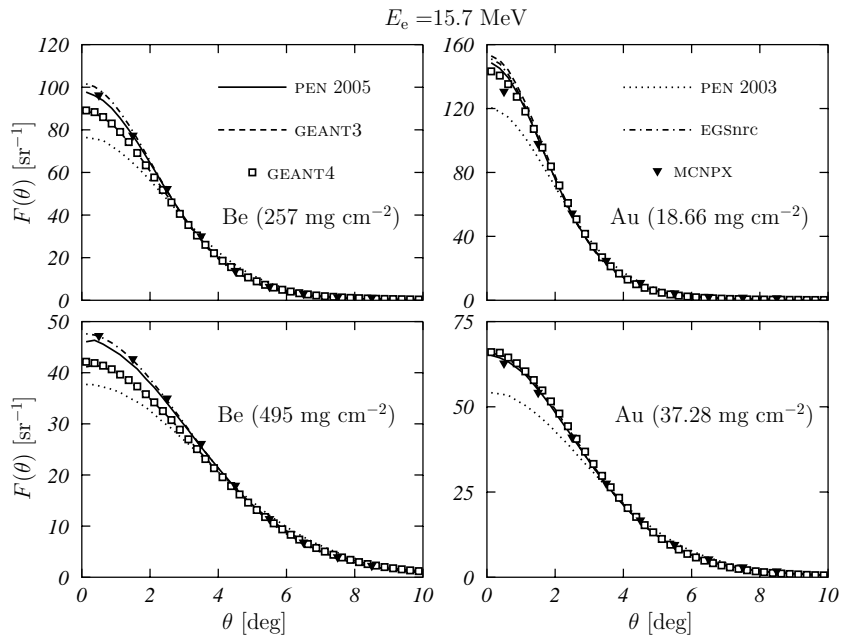}}
\end{center}
\caption{\small Angular distributions obtained with the various MC
codes for the four foils of the experiment of Hanson {\it et al.}
\cite{han51}. Statistical uncertainties (1$\sigma$) are smaller than
2\% and have not been plotted.
\label{fig:comp-CODES-HA}
}
\end{figure}

\subsection{Comparison of the simulations}

To finish the discussion we compare now the angular distributions
obtained with the different simulation codes.

In Fig. \ref{fig:comp-CODES-HA} we show these distributions for the
four foils of the experiment of Hanson {\it et al.} \cite{han51}. As we
can see, the behaviour observed is different in the case of the Be
foils (left panels) than in the Au ones (right panels). For Be, the
calculations performed with \EG \/ (dashed-dotted curves) show the
maximum values $F(0)$, and are in rather good agreement with the
results obtained with \PENF \/ (solid curves) and \MC \/ (solid
triangles). The \GT (dashed curves) \/ and \GF \/ (open squares)
results overlap for all values of $\theta$ and, finally, the \PENT \/
(dotted curves) are clearly below the other calculations at small
angles. The relative differences between the maximum and minimum
values of $F(0)$ are 33\% and 26.3\% for the thin and the ticker
foils, respectively. 

\begin{figure}[t]
\begin{center}
\parbox[c]{15cm}
{\includegraphics[scale=0.6]{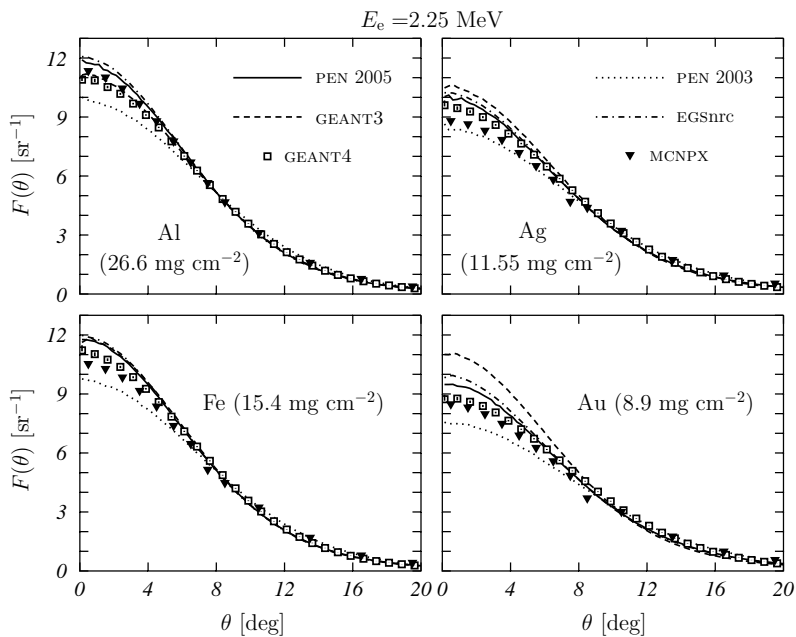}}
\end{center}
\vspace*{-.5cm}
\caption{\small Same as in Fig. \ref{fig:comp-CODES-HA} but for the
Al, Fe, Ag and Au foils in the experiment of Kulchitsky and Latyshev
\cite{kul42}.
\label{fig:comp-CODES-KU}
}
\end{figure}

In the case of the Au foils (see right panels), the situation of the
\PENT \/ results is the same, but the other calculations are all of
them grouped. This is particularly clear in the case of the thicker
foil (lower right panel).

The results obtained for the different foils considered by Kulchitsky
and Latyshev \cite{kul42} show up a different behaviour. In
Fig. \ref{fig:comp-CODES-KU} we have plotted the angular distributions
corresponding to the Al, Fe, Ag and Au foils. Similar results are
found for the other four foils studied in this experiment.  \PENT \/
provides the smaller $F(0)$ values in all cases except for Mo, in
which \MC \/ is the code showing the minimum. \EG \/ shows the larger
values for Al, Fe and Cu and \GT \/ for the remaining foils. As we can
see, the results appear to be spread for forward angles and the
spreading increases with the atomic number of the material. The
relative differences between the maximum and minimum values of $F(0)$
range between 17.3\% for Cu to 45.9\% for Pb, being $\sim 20\%$ for Al
and Fe. In all cases analyzed, \PENF \/ and \EG \/ are in good
agreement.

To finish, it is worth to point out the situation observed for the Au
foils. As we can see (lower right panel in
Fig. \ref{fig:comp-CODES-KU}) the various codes produce rather
different angular distributions for $\theta < 8$ degrees. This
contrasts with the overlapping shown by the results corresponding to
the Au foils in Fig. \ref{fig:comp-CODES-HA} (see right panels.)

\section{Conclusions}

In this work the experimental data available for multiple scattering
by thin material foils have compared to the simulation results
obtained with the MC codes \EG, \MC, \GT, \GF, \PENT \/ and \PENF.

Simulations performed with various older versions of \GF \/ show up
non-negligible differences in the angular distributions with respect
to the version 8.0, differences which are bigger the lower the $Z$ of
the material is.

Different simulations done with \PEN \/ have indicated large
differences in the angular distributions between the versions 2003 and
2005, though the mean values $\langle \mu \rangle$ and $\langle \mu^2
\rangle$ obtained with both versions are very similar, as expected.
In addition, we have found discrepancies between the ``safe'' and
``detailed'' simulations which are related to the number of
interactions suffered by the electrons in the foil and which can be
largely reduced by an adequate selection of the tracking parameters.

We have compared the results of our simulations with the experimental
data of Kulchitsky and Latyshev \cite{kul42} by means of the
characteristic angle $\theta_{1/e}$. We have found that both \EG \/
and \PENF \/ show differences with respect to $\theta^{\rm exp}_{1/e}$
which are within a range of 3\% for all foils. However, the better
description of the experimental data is provided, globally, by \GF.

The comparison with the results of Hanson {\it et al.} \cite{han51} shows
that \EG, \GT, \GF \/ and \PENF \/ give a good description of the data
in the case of the Au foils. For the Be foils, the better agreement
with the experiment occurs for \EG, \MC \/ and \PENF.

All codes (except \PENT) show a good agreement for the two Au foils in
the experiment of Hanson {\it et al.} For the Be foils, the maximum relative
differences between the codes at $\theta =0$ are larger than 25\%. In
the case of the foils in the experiment of Kulchitsky and Latyshev,
these maximum relative differences grow with $Z$, reaching $\sim 50\%$
in the case of the Pb foil. In general, \EG \/ and \PENF \/ are in
rather good agreement for all cases studied.

The effects observed in these thin material foils appear to be
non-negligible. Differences between the various MC codes are relevant
and the propagation of them to other situations should be investigated
in detail. One of these situations concerns the simulation of electron
linear accelerators, where different thin foils are present. The
combined effect of all of them could modify the conclusions found in
previous works in which various codes have been intercompared
(e.g. in Ref. \cite{sem01}.)

On the other hand it should be interesting to perform new experiments
with different materials in order to permit a complete test of the
multiple scattering theories used in the MC simulation codes of the
radiation transport.

\section{ACKNOWLEDGMENTS}
We want to acknowledge useful discussions with F. Salvat,
J.M. Fern\'andez-Varea, M.A. Coca and D. Guirado. This work has been
supported in part by the Junta de Andaluc\'{\i}a (FQM0220) and by the
Ministerio de Educaci\'on y Ciencia of Spain (FIS2005-03577).

\end{document}